# Deriving Ontologies from XML Schema

Ivan Bedini*, Georges Gardarin**
Benjamin Nguyen**

*Orange Labs, Site of Caen - 14000, France
ivan.bedini@orange-ftgroup.com; http://www.orange.com

**Prism, Université de Versailles St-Quentin, 78035 Versailles Cedex 035
<prenom>.<nom>@prism.uvsq.fr; http://www.prism.uvsq.fr

**Abstract.** In this paper, we present a method and a tool for deriving a skeleton of an ontology from XML schema files. We first recall what an is ontology and its relationships with XML schemas. Next, we focus on ontology building methodology and associated tool requirements. Then, we introduce Janus, a tool for building an ontology from various XML schemas in a given domain. We summarize the main features of Janus and illustrate its functionalities through a simple example. Finally, we compare our approach to other existing ontology building tools.

## 1 Introduction

Ontologies appear as useful building blocks in several domains including the Semantic Web, data source integration, data visualization and zooming, text indexing and mining, etc. In Data Warehouse and On Line Analysis, ontologies are particularly useful for integrating multiple data sources and for zooming on data cube dimensions. They provide a deep understanding of data by composing and transforming it automatically. Though helpful and trendy, ontologies are still hard to build, to maintain and make evolve, notably when large (e.g., more than thousands of concepts).

The "nuts and bolts" of ontologies are concepts and attributes with *is-a* and *part of* relationships. In that respect, ontologies are similar to object models of specific domains. XML schemas also have similarities with object models. While there exist many XML schemas in certain domains (eg., B2B e-commerce are at standardization level), there are only very few *ontologies*. In this paper, we propose a methodology and a tool for semi-automatic derivation of ontologies from XML schemas. We present the main features of our tool called Janus and illustrate it through a simple example. Janus is unique in the sense that it mixes several technologies (language analysis, text mining, schema validation, graphical representation) into an incremental methodology of ontology construction and evolution. We further propose a classification of ontology building tools and their comparison in terms of functionalities.

The rest of this paper is organized in four main sections: i) ontology definition and requirements for building tools; ii) overview of our Janus ontology skeleton building tool; iii) presentation of other tools; and iv) comparisons of tools.

## 2  Ontologies : Definition and Requirements

### 2.1  Definition

An ontology is a common abstract and simplified view of an application domain, e.g., finance, tourism, medicine, transport, B2B, wine, etc. It defines the concepts, properties and relationships of the field. It also aims at describing the objects and their relations according to a common point of view to which all participants agree. Concepts are often assimilated to object classes and properties to object facets. Among the facets, some are mono-valued while others are multi-valued. A mono-valued property corresponds to a function having for domain a class and for image a data type or a class.

Formally, according to (Amann *et al.* 2000), an ontology is a triple (C, S, ISA) where:

(i) C is a set of classes $c_1, c_2, \ldots c_m$ ;

(ii) S is a set of properties $s_1, s_2, \ldots s_3$ ; a data property (integer, real, string, date, etc.) is an attribute denoted by a name and having value in a base type; an object property represents a binary relation role characterized by a name and a target class ;

(iii) ISA is a set of inheritance links defined between classes.

This definition can be extended with instances of classes, i.e., objects, and with constraints between classes and between properties.

### 2.2  Ontology Construction Requirements

By building an ontology, we aim at constituting a knowledge representation integrating all the information and the suitable rules of a domain. This is hardly possible without a staged approach: an ontology must be developed progressively without disturbing existing applications when a new version is adopted; a basic ontology must be *enriched* by successive extensions. The construction technique must maintain discovered extensions and use them to identify quickly similarities between new and old concepts. These characteristics, together with the definition lead to the following ontology construction tool requirements: efficient memorization of concepts/relationships for reuse with clear distinction of polyseme, good support for dynamic evolution and resourceful automation of production.

The foster is an important issue because as pointed out in (Ehrig *et al.* 2004) and (Sabou *et al.* 2006), time performance for computing data integration is a key problem when mapping ontologies. Thus, a solution able to store the best known concepts for a domain, should generate better reusability of components by machines applying reasoning algorithms.

Also to maintain and enrich an ontology as automatically as possible is a fundamental requirement. Even in a specific field, the concepts handled by the applications can be numerous and the quantity of information which we wish to maintain for each concept is vast. Solely relying on human management becomes impossible: for example, consider a corpus made up of a thousand of files with concepts per hundreds in each file: this is a task which can not be completely manually.

### 2.3  Ontology and Schema

XML schemas and ontologies in a given domains are somehow related. In general, schemas are built in a domain before ontologies. Consider for example the B2B domain: there exist hundreds of schemas to encode exchanged data but not many ontologies. To benefit from preexisting schemas, we propose a method and a tool to derive an ontology or at least an enriched taxonomy (i.e., a concept hierarchy with concepts properties and main relationships) from a set of XML schemas.

Let us give two examples of schemas that can be used in the wine domain. The problem is how to derive the wine-tasting ontology from these schemas, or at least something resembling a subset of this ontology. From this first version, knowledge could be added (possibly derived from new schemas) or updated (by an expert) to derive a satisfactory ontology.

**Example 1 – Example of Schemas in the wine-tasting domain**

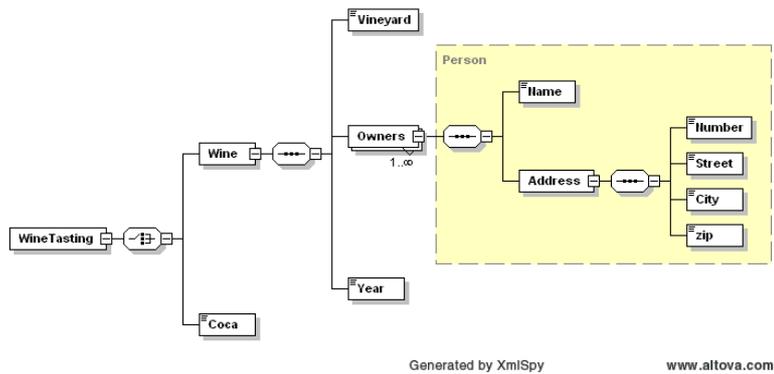

FIG. **1-** *Wine Tasting XSD*

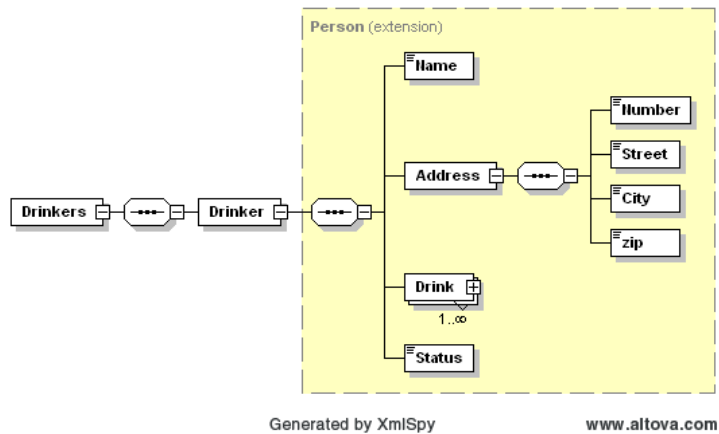

FIG. **2-** *Wine Drinkers XSD*

## 3  Deriving Ontologies from XML Schemas

### 3.1  Methodology

The aim of this paper is to provide a general view of the automation aspect of the ontology generation; thus, in the rest of this section, we provide some elements that we consider crucial in order to achieve this result.

Several methodologies for building ontologies exist, such as OTK (Sure *et al.* 2004), METHONTOLOGY (Corcho *et al.* 2003) or DILIGENT (Vrandecic *et al* 2005), but they target ontology engineers and not machines. We do not develop here a new methodology yet, but we define the automatic ontology generation life cycle as a process composed of five main steps necessary to achieve our goal. These steps represent the main tasks of the process for building ontologies starting from an existing corpus, like XML schemas. We do not focus on what techniques are available for each task, but mainly describe what we expect from each task. The whole process is depicted in FIG. 3. The five steps are:

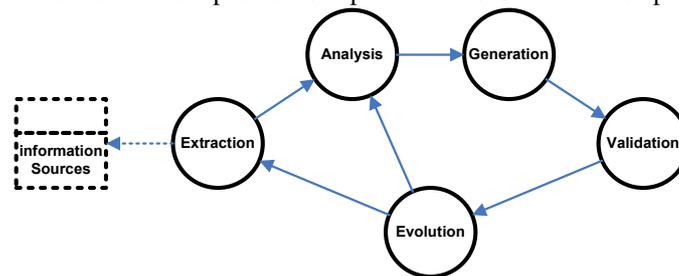

FIG. 3 – *Automatic ontology generation process*

**1. Extraction:** This step deals with the acquisition of information needed to generate the ontology (concepts, attributes, relationships and axioms) starting from an existing corpus. Input resources can be of many types: structured, semi-structured or unstructured. Techniques for information retrieval and extraction can be of different types, such as: NLP (Natural Language Processing), word clustering, machine learning, semantic, morphological, lexical approaches, and more often a combination of them.

**2. Analysis:** This step focuses on the matching of retrieved information and/or alignment of two or more existing ontologies, depending on the use case. This step requires techniques already used in the first stage, as morphological and lexical analysis of labels; a semantic analysis to detect synonyms, homonyms and other relations of this type; an analysis of concept structures to find hierarchical relationships and identify common attributes; techniques based on reasoning to detect inconsistencies and induced relations.

**3. Generation:** This step deals with ontology merging, if appropriate, and the production of a first version of the target ontology based on the tool formal meta-model, i.e., in a universal language interpretable by other applications, such as OWL and RDF/S.

**4. Validation:** All previous steps may introduce wrong concepts or relationships, thus a validation phase is needed. Conversely, a validation task can be introduced at the end of each previous step. Validation is often done by hand, but can sometimes be automated.

**5. Evolution:** An ontology is not a static description of a domain, but with the evolution of applications, in quality and number, the ontology may also require some changes. Some concepts, as well as properties, relationships, and other parameters, can be added or modified. Incremental construction should be supported efficiently and could integrate a new step of information extraction, if new resources are not yet integrated in ontology format. It could also directly provide new matches and alignments by integrating a new analysis step.

## 3.2 Deriving Concepts and Properties

Unlike simple text documents, XML documents provide likely annotated text with important information about objects and their structures, thus concepts for the ontology to build.

Considering the examples presented in **FIG.** 1, it is pretty simple to imagine equivalences between OWL classes and XSD elements, like *Wine* or *Drinker*; between OWL data-type properties and XSD element pointing to a simple type, like *Year* or *Name*. Also we can retrieve information about relationships like *IS A* (e.g.: an *Owner* is a *Person*), thus provide a classification of concepts, etc….

We can formalize ontology learning concepts and properties from XSD for as follows:

*Definition 1*. Given a set of XSD files *X* as input source, it is possible to retrieve a complete set of concepts and properties *O* by a surjective mapping $m^1$ (based on XML mining techniques), $m : X \rightarrow O$, that we call **domain conceptualization**.

*Definition 2*. A **concept** is the basic element of *O* and is defined as a quadruple c = <L, Hc, Rc, I> where:

- the label *L* is a common word (simple or compound) that best represents the concept. It is selected from a close set of names extracted from *X* that can be associated to the concept (e.g.: *owner* and *person*).
- *Hc* is the set of structural relationship, which correspond to the subsumption hierarchy (fundamentally *IS A* relationship and *propertyOf*).
- *Rc* is the set of relations partitioned to two subsets. One is the set of all assertions in which the relation is a semantic relation (like synonymy) and the other one is the set of all assertions in which the relation is a non-semantic relation (like usage for common label *l* abbreviations, or proved similarities such as the correspondence *Person.Name => Person.FirstName + Person.LastName*).
- *I* is the set of originating instances of a concept, a link to the source.

Depending on domain usages it is important to know if a concept represents a detail of the corpus source or a main concept. For example when looking for characteristic concepts of the wine ontology, we can state that *wine* and *person* best represent the domain than *status* or *address,* because in this context the latter are properties. More precisely we are able to specialize concepts in order to obtain a more grained refinement of concept roles as follows:

*Definition 3*. Let *C* represent a set of concepts called the set of concept **classes**, $C = \{c1, ..., cm\}$ a finite subset of *O*. A concept is considered to be a class, thus element of *C* if it has more than one property. We say that a concept $c \in C$ if $P(c)= \{c1, ..., cm\}$, for $m > 1$, and of course $P(c) <> 0$. $P(c)$ is called the set of properties for a given concept *c* and *c* is a class.

*Definition 4*. Let *P* represent a set of concepts called the set of concepts **properties** a finite subset of *O*. A concept *cp* is considered a concept property if exists at least one super class of which it is a property. A concept $cp \in P$ if $\exists c \in C \mid cp \in P(c)$

*Definition 5*. Let *D* represent a set of concepts called the set of **printable** concepts, also called **data-type** concepts, a finite subset of *O*. A concept *cdt* is considered a data-type concept if it has no properties and it is directly related to a printable type (as defined by the XML Schema built-in Datatype list[2]).

## 3.3 Deriving Some Rules

As seen in the previous section, it is possible to retrieve information about concepts from XSD. For this it is necessary to provide a set of rules that we can summarize with the following principles:

---

[1] A mapping from set A onto B is called surjective (or 'onto') if every member of B is the image of at least one member of A. ➔ $f : A \rightarrow B$ *is surjective if* $\forall b \in B ( \exists a \in A (f(a)=b))$

- XSD complex types with complex content (i.e., a combination of attributes and sequence of elements like for *Wine*) produces concepts classes, otherwise printable types;
- XSD elements can assume different facets as simple properties if they point to a simple type or a printable structure (like *Name* and *Street*), as classes if they declare a complex content (like *Drinker*), as a specialization of a class or a role, if they are named elements with declared type (like *Owners*);

Starting from these macro rules it is possible to build a complete list of rules for the surjective mapping (see Definition 1) as shown in **TAB. 1**. The first column lists main XSD structures as defined by XML Schema primer recommendations, while the Mapping column provides the corresponding domain conceptualization.

| XSD Structure | Mapping to *O* |
| --- | --- |
| xs:complexType | Concept class |
| xs:simpleType | Concept datatype |
| xs:extension et xs:restriction | Datatype property and *is a* relationship |
| Xs:union | ComplexType properties |
| xs:any | Datatype property of the correspondent concept |
| xs:simpleContent and xs:complexType with declared xs:simpleContent | Concept datatype |
| Element with attribute "ref" to xs:complexType | Concept class with *propertyOf* relationship |
| Named xs:element with attribute "type" | Concept class with *Is a* relationship |
| Named xs:element | Concept class without attribute type |
| xs:minOccurs, xs:maxOccurs | Respective cardinalities |
| xs:sequence, xsd:all | Concept properties |
| Attributes of xs:element and xs:compleType | Concept properties |
| xs:choice | Disjointness concepts |

TAB. 1 - *XSD Mining information extraction and correspondent mapping*

Working on XSD we just target TBox statements, but anyway we are also able to define rules for detecting subsumption, equivalence, disjunction and classification relationships between concepts. For example we can observe that the concept *Drinker* subsumes *Person* because it is less general (in fact XSD file declares that it is an extension). Also when tasting, *Coca* and *Wine* are disjoint classes (within the XSD the *WineTasting* element proposes a choice between these two classes). We can classify concepts as classes or properties and look for equivalences like *Owner* and *Person*.

When concepts are extracted following rules in **TAB. 1**, it is common to find different formalizations of the same set of concepts from different sources. Then, we need to define other rules to match similar concepts. Similarity between concepts can be of three types. The first one uses the semantics of labels (synonyms, lexical, common terms for compound words like *PostalAddress* or *GeographicalAddress*), the second is related to properties (like two concepts with different label but exactly the same consistent set of properties) and the last looks for common relationships between concepts, considered as the concept context.

To conclude we can state that the derivation of all these rules lets us to obtain a consistent mapping producing the conceptualisation of a domain, a skeleton of the domain ontology.

### 3.4 The Janus Prototype

Despite the large adoption of XML files and schemas, only few tools provide advanced ontology learning from this format. They are mostly limited to language translations (like XML conversion to OWL), and mining of one file at a time. Real practices, like in the B2B domain, show that each workgroup often defines several XSD instances for related domains

I. Bedini et al.

(Bedini *et al.* 2008). This precludes the usage of existing tools to such use cases. This is the motivation that brought us to develop a new tool.

The main idea of our approach is: i) for each file to capture concepts from words contained in XML tags; ii) to get relationships from XML structures; iii) to match concepts similarities and; iv) to merge concepts to produce a general view of input sources. In fact, this approach can be applied to different kinds of sources (like UML, WSDL, OWL…) because the conceptualization is general enough, but currently we focus on XML schemas.

### 3.4.1 Architecture

Our prototype, called Janus, follows the generation process defined in section 3.1 and implements rules defined above in order to generate an ontology from XML heterogeneous corpora as automatically as possible.

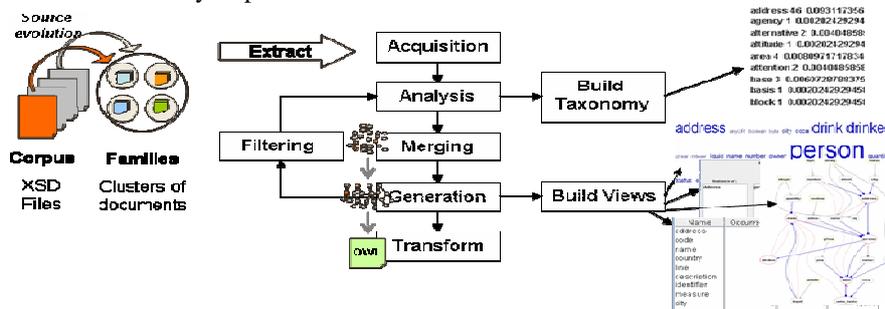

FIG. 4 – *Janus overall architecture*

**FIG. 4** shows the overall architecture of Janus which is composed of three main steps. The **Extraction** and **Acquisition** tasks provide the knowledge retrieval needed to generate the ontology (concepts, properties and relationships). Implemented techniques for this task are a combination of different types, such as: NLP (Natural Language Process) for morphological and lexical analysis, association mining for calculating term frequencies and association rules, semantics to detect synonymy, and clustering for grouping semantic and structural similar concepts. We refer to the adaptation of these techniques as XML Mining. The **Analysis** step focuses on the matching of retrieved information and alignment of concepts issued from different sources. This step requires techniques already used in the first stage, as syntax and semantic measures, to establish the best similarities; it also requires an analysis of concept structures to determine hierarchical relationships and identify common properties. The last step is composed of **Merging** and **Generation** tasks and looks for concepts with evident affinities (e.g., concept fully included into another) to merge them. It transforms the metamodel used by the tool into a semantic network that can be described in RDFS or OWL. The tool can derive from the network useful views provided to users, which can also step into the process to parameterize thresholds for refining results.

### 3.4.2 Application to the Wine Example

The example above does not show great complexity due to the fact that only one description is provided for each concept and that semantics are well formed (only real words are used). Nevertheless the tool is able to catch all concepts and relationships, in respect to humans, just one relationship is quite wrong. It is the plural form, provided by drinkers, this

EDA 2008

because the tool does not make difference between singular and plural yet. The table below resumes concepts discovered and their relationships.

| Classes | Object Properties of related classes | Object Datatypes of related Properties | Relationships |
|---|---|---|---|
| 6 (wine_taste, wine person drinker drink address) | 12 (quantity, vineyard, year, zip, status, city, coca, street, boolean, liquid, name, owner) | 7 (string anyURI gYear token integer byte number) | Owner IS A Person Drinker IS A Person Coca DISJOINT Wine |

TAB. 2 – *Tasting Wine XSD Mining concept extraction*

# 4  Overview of Ontology Building Tools

A brief glance at current solutions of automatic ontology building systems, sometimes referred to as Ontology Learning, shows many problems still need solving. In this section, we cover a large spectrum of existing ontology building tools, starting with general articles.

(Mehrnoush, Abdollahzadeh, 2003) present a complete framework that classifies software and techniques for building ontologies in six main categories (called *dimensions*). It is a detailed and interesting classification, but it focuses only on the learning method. (Buitelaar *et al.*, 2005) provide a comprehensive tutorial on learning ontology from text, which is really useful, but the considered corpus source does not fit our use case. (Euzenat *et al.*, 2004) provide a detailed analysis of technical alignment for ontologies and a state of the art on existing tools, probably the best known matching and alignment software, but they concentrate on the one task of aligning two existing ontologies, without investigating other steps in the generation process, such as information extraction or the problem of multiple merging. (Castano *et al.,* 2007) are also limited to two existing ontologies.

Despite these documents, it remains difficult to clearly understand who makes what and why. This is because frameworks designed for evaluation and analysis tend to organize methods according to adopted technologies. Our approach wants to facilitate the understanding of what a method does within the ontology generation life cycle. The process defined in Section 3.1 constitutes our base framework for evaluating software and experiments presented in this Section. It provides us elements to evaluate which part of the process can be automated and how, as well as which techniques are more appropriate, and if they still require human intervention, thus further research. In this paper, we complement the documents above, and overlap on tools that are closest to our interests.

## 4.1  Ontology Generator Classification

Ontology generation is mainly hand-made by domain experts, but this is of no interest to us. In this paper we group experiences and software in four main categories as follow:

**- Conversion or translation** for applications that make the hypothesis that an ontology is already well defined by someone or somewhere. The ontology format representation is wider than other common knowledge representations, such as XML or UML. Software is produced to compute this transformation. Experiments show that this approach presents a high degree of automation, mainly because it does not address the whole problem of ontology generation.

**- Mining based** for applications implementing mining techniques in order to retrieve enough information to generate an ontology. Most experiences are focused on unstructured sources, like text documents and implement Natural Language Processing (NLP) techniques. These experiences tell us that recovering structured concepts from unstructured documents still requires human assistance and that mining techniques from natural text can be used only in complement with other existing structured knowledge representations techniques.

   **- External knowledge based** for applications that build or enrich a domain ontology by using an external resource. This category sometimes overlaps the mining based because techniques applied to retrieve information can be the same, nevertheless we classify here experiences with an approach closer to the integration of external dictionaries, existing ontology or from a more general knowledge resource, like WordNet (Miller, 1995) or the Web.

   **- Frameworks** for applications with an approach based on integrating different modules.

As always when creating a classification, the border is not well defined, therefore we classify works with respect to their automation approach rather than with regards to the techniques they implement. In fact we support the thesis that there is not a single technique to develop, but that only an appropriate mix of techniques can bring us to our goal.

We will now describe the software and experiences, using our classification.

## 4.2 Conversion or Translation

The methods presented here assume that the source of the transformation is a complete and well-defined ontology in XML or UML format.

(Bohring, Auer, 2005) have developed a tool that converts given XML files to OWL It is based on the idea that items specified in the XSD file can be converted to ontology's classes, attributes and so on. Technically they have developed four XSLT instances to transform XML files to OWL, without any other intervention on semantics and structures during the transformation. This method has been applied to the Ontowiki platform (Auer *et al.*, 2006).
Similarly, (Gasevic *et al.*, 2004) propose using UML profiles to overcome UML's limitations, also using XSLT instances to convert to OWL.

(Bertrand*, et al.*, 2006) propose a semi-automatic process, starting from a UML class diagram representation of the ontology domain. A human selected part of the UML model is then transformed into ODM (Ontology Definition Metamodel) as pivot model before automatically generating an RDFS file.

Within the PICSEL project, (Giraldo, Reynaud, 2002) have developed a semi-automatic ontology generation software for the tourism industry domain extracting information contained in DTD files. This experience is interesting because it goes further, than the XML to OWL transformation seen previously, and shows that *tags and structure of XML files have sufficient information to produce an ontology*. Human intervention is needed to detect abbreviations or false positives during the ontology validation task. This experience is really close to our use case but is limited to the sole domain of tourism; therefore the detection of relevant concepts does not produce conflicts between different representations.

## 4.3 Mining Based Approaches

The general drawback of these methods is the use of a more general ontology to define concepts for the targeted ontology (for instance WordNet). Our feeling is that these approaches should be used in complement of the automatic ontology generation process.

(Biebow, Szulman, 1999) present the TERMINAE method and tool for building ontological models from text. The method is divided into 4 stages, corpus selection and organisation; linguistic analysis with the help of several NLP tools, such as LEXTER (Bourigault, 1996); normalization according to some structuring principles and criteria; formalization and validation. An expert is called to select the most important notions (concepts) for the targeted ontology from the list of *candidate terms extracted by the tool* and to provide a definition of the meaning of each term in natural language. These terms may or may not be inserted.

(Lonsdale *et al.*, 2002) propose a process to generate domain ontologies from text documents called SALT. Their methodology requires the use of three types of knowledge sources: one is a more general and well defined ontology for the domain, a dictionary or any external resource to discover lexical and structural relationships between terms and a consistent set of training text documents. Using these elements they are able to automate the creation of a new sub-ontology of the more general ontology. User intervention is required at the end of the process to remove false positives. The authors state that with a large set of training documents their solution can achieve good results.

Similar to (Lonsdale *et al.*, 2002), (Kitz *et al.*, 2000) describe a generic approach for the creation of an ontology for a domain based on a source with multiple entries which are: a generic ontology to generate the main structure; a dictionary containing generic terms close to the domain; and a textual corpus specific to the area to clean the ontology from wrong concepts. This approach combines several input sources, for greater generality and better reliability of the result. However, a manual check of the ontology must still be performed.

(Hu, Liu, 2004) have developed an automatic generation based on an analysis of a set of texts followed by the use of WordNet. The analysis of the corpus retrieves words as concepts. These words are then searched in WordNet to find the concepts. The ontology generation seems to be one of the most automated, but no details of how the terms are extracted from the body nor any qualitative assessment of the work are provided.

## 4.4 External Knowledge Retrieval

Also based on WordNet, (Moldova, Girju, 2000) expose a similar method but with the difference that if a word is not found in WordNet then a *supplementary module will look for it over the Internet*. Then linguistic and mining techniques extract new "concepts" to be added to the ontology. User intervention is necessary here to avoid incongruous concepts.

(Aguirre *et al.*, 2000) have developed a strategy to enrich existing ontologies using the Web to acquire new information. The method takes as input a word which one wants to "improve" the knowledge of. WordNet is questioned about this word, and the different meanings are used to generate queries for the web. For each query, that constitutes a "group", different search engines are queried. Terms frequencies are then calculated and compared with each group, and of course the winning group, (i.e. sense), for the concept is the one with the highest frequencies. In addition a *statistical analysis* is performed on the result, in order to estimate the *most common meaning* of the concept. This method alone can not be adopted to build ontologies, but it has the merit to be able to iterate with an external knowledge base to provide further information that may be used for the validation task of an ontology in absence of human intervention.

(Cho *et al.* 2006) present the problem of proximity between two ontologies as a choice between alignment and merging. The first case is limited to establishing links between ontologies while the second creates a single, new ontology. With their experience they directly merge two ontologies based on WordNet. For this they use two approaches in their method that they call the horizontal approach and the vertical approach. The horizontal approach first checks fall the relationships between concepts of the "same level" in the two ontologies and merges or ties them as defined by WordNet, while the vertical approach completes the merg-ing operation for concepts with "different levels", but belonging to the same branch of the tree. In this case they fill the resulting ontology with concepts from both ontologies and do not make a choice. A similarity measure is calculated in order to define the hierarchy between these concepts in the resulting tree. This method, while not providing an adequate solution to automation, does provide a *purely semantic approach* to the merging solution.

### 4.5 Frameworks

SymOntoX (Missikoff, Taglino, 2003) is an OMS (Ontology Management System) specialized in the e-business domain, which provides an editor, a mediator and a versioning management system. The creation of the ontology is mainly done by an expert using the editor, but the framework contains a first step towards an easier generation: it contains high-level *predefined concepts* (such as Business Process, Business Object, Business Actor, etc.), as well as *different modules* used for ontology mapping and alignment to simplify the work of the expert. Ontology generation is not automatic, but merely assisted.

Protégé (Noy *et al.*, 2000) is a free, open source, platform to design ontologies. It is supported by a strong community and experience shows that Protégé is one of the most widely used platforms for ontology development and training. This software has an *extensible architecture* which makes it possible to integrate plug-ins. Some of these modules are interesting and relevant in our case, like those from the PROMPT Suite (Noy, Musen, 2003). They automate, or at least assist, in the mapping, merging and managing of versions and changes. Also the related project Protégé-OWL offers a Java API to manage OWL and RDF formats.

The glue between these pieces of software still remains human, yet program modules and libraries provide a fundamental basis for developing the automation of ontology generation.

(Maedche, Staab, 2003) are contributors of several interesting initiatives within the ontology design field as well as the automation of this process, like MAFRA Framework, Text-To-Onto and KAON. In this paper we focus on their framework for ontology learning.

They propose a process that includes five steps: import, extraction, pruning, refinement, and evaluation. This approach offers their framework a flexible architecture that consists of many extensible parts, such as: a component to manage different input resources, capable of providing information extraction from a large variety of formats (UML, XML, database schema, documents text and web); a library of algorithms for acquiring and analyzing ontology concepts; a graphical interface that allows users to modify the generated ontology, but also to choose which algorithms to apply and treatments to perform.

They bring together many algorithms and methods for ontology learning. Despite their framework not allowing a completely automatic generation process, they are the only people to propose a *learning process* close to a methodology of automatic ontology generation.

(Raghunathan, 2003) introduces the system LOGS (Lightweight universal Ontology Generation and operating architectureS). He states that generating ontology automatically from text documents is still an open question. Their system is developed with a modular architecture that integrates the core functionality that can be expected by automatic ontology building software. It consists of the following modules: document source parser, NLP engine, analyser, ontology engine, interface, integrator, ontological database and dictionary. It also contains other modules able to crawl an intranet, refine the process of ontology design and a module implementing trial and error iterative analysis of related texts to find patterns. No qualitative analysis is provided, but the authors argue that they obtained significant results. Unfortunately this software seems to have not met great consensus within the community.

Deriving Ontologies from XML Schema

| | Extraction | Analysis | Generation | Validation | Evolution |
|---|---|---|---|---|---|
| **Generating an ontology from an annotated business model** | - Human | - | C – No merging. Direct transformation using XSLT files. | - Human, upstream to the generation | - |
| **XML2OWL** | B – Static table of correspondences | - | C – No merging. Direct transformation using XSLT files. | - Human, upstream to the generation | - |
| **UML2OWL** | B | - | C – No merging. Direct transformation using XSLT files. | - Human, upstream to the generation | - |
| **Semi-automatic Ontology Building from DTDs** | C – automatic extraction from DTD Sources | B – structure analysis without alignment | C – No standard ontology representation | - Human | - |
| **Learning OWL ontologies from free texts** | C – Text sources. NLP techniques. WordNet as resource dictionary/ontology | - | C – OWL format | - | - |
| **TERMINAE** | C – Text sources. NLP techniques | B – Concept relationships analysis | C – No standard ontology representation | - Human | - |
| **SALT** | D – Text sources. NLP techniques. Multi entries. | C – Similarity analysis of concepts | B – No standard ontology representation | B –Limited human intervention | - |
| **A new Method for Ontology Merging based on Concept using WordNet** | - | B | C – Automatic merging. No standard ontology representation. | - | - |
| **Enriching Very Large Ontologies Using the WWW** | C – Enrich existing ontology | - | C | - | - |
| **Domain-Specific Knowledge Acquisition and Classification Using WordNet** | C – Main concept defined by a domain expert. | B – Grammatical analysis of text | C | - Human | - |
| **A Method for Semi-Automatic Ontology Acquisition from a Corporate Intranet** | C – NLP techniques. Multi entries source. | B – Meaning analysis of concepts | B | B – User required for undecidabe cases | B – Cyclic approach can manage evolutions |
| **SymOntoX** | - | C – Matching analysis | B - Provide some predefined concepts. | - Human | B – Manage versions, but still human. |
| **Protégé (Mainly from plug-in)** | B – extraction from Relational DB and some XML format | D – Matching and Alignment analysis. | B – Assisted merging. Export in several ontology formats. | - Human | C – Ontology evolution detection |
| **LOGS** | C – Text source analysis. NLP engine. Morphological and semantic analysis. Machine learning approach for rules. | C – Similarity based on concepts and relationships analysis. | C – Different format. Internal ontology structure based on a lattice. | B – Validation at the end of each module | - |
| **Ontology Learning** | D – Extraction from several formats (XML, UML, OWL, RDF, text…). NLP, Semantic and lexical analysis. Multi entries source. | C – Libraries for clustering, formal concept analysis and associations rules | C - OWL and RDF/S | B - Assisted | - |

TAB. 3 – *Comparative analysis of methods*

## 5 Comparative Analysis and Discussion

Section 4 presents only a subset of all the works we have studied, but we have tried to cover all types of approaches. In this section, we provide a comparative analysis of methods following the five steps composing the automatic ontology generation process. We present the strengths and weaknesses of experiments implementing at least one step of this process.

The degree of automation can not be measured exactly. Moreover, qualitative results were not always available. When conducting this assessment only three tools presented in this paper were both freely available and able to process XML Schema files (as required by our use case), and therefore specifically tested by us. These are Protégé, XML2OWL (Bohring, Auer 2005) and MAFRA (Maedche, *et al.*, 2002). Despite this lack of availability, the purpose of this study is mainly theoretical, thus information obtained by public material was enough to perform a qualitative evaluation.

Values are assigned to each step according to the following criteria:

A– when step is not developed (marked by a – symbol)
B – for solutions using a semi-automatic approach
C – for solutions where human intervention is optional
D – for solutions that are, a priori, completely automatic

We draw the following conclusions from the analysis of Table 3: Information extraction can reach good results. The most studied input corpora are for text documents, a lot of information can be reached from this type of corpus source. Methods based on this type of resource have the advantage to have a lot of resources, that can be found over Internet or an Intranet, and that several tools for NLP and mining are available. Nevertheless they require a most important human validation task and are preferred for defining a high level definition of concepts. Structures, like classes, attributes and relationships, are mostly provided by other external resources. Thus mining directly structured documents can reach better results with less validation, but not so much methods deepening study this approach.

To this end WordNet surely deserves some special attention because we can observe that it is an essential resource for the automation process. In fact it plays different roles. The first is that of an electronic dictionary and thesaurus, which is fundamental. The other is that of a reference ontology, mainly by using its sibling or hierarchical terms discovery, with relationships like hyponym, meronym, holonym and hyperonym. WordNet has the drawback of being too generic and not adapted to specific domain ontology development. Even so, it remains an important module to further be developed.

Matching and alignment modules are the most challenging tasks but, as told in (Euzenat *et al.*, 2006), they are growing and methods and techniques in the future should achieve valuable results. Such modules should be available as shared libraries.

Merging, which is strictly related to alignment, is currently implemented with two input ontologies. Multi ontology alignment seems to be an open question yet to be investigated in detail. This point could be resolved with consecutives merges, but it appears that the final ontology can be different depending on the sequence in which the ontologies are merged.

Validation is almost always human and only automatic consistency checking has been implemented. The only solution to improve it, is to limit its range, thus: adopting a bottom-up approach, which has shown better results; to use successive refinements and reasoners, in order to guarantee consistency in the resulting ontology and; by querying external resources like Watson (d'Aquin *et al.*, 2007) rather than the Web directly, that provides the advantage of returning structured information, which is more suitable for machine interpretation.

Evolution management is still rare. Some methods manage versions and other go further and provide automatic detection of changes. But in reality what we are really looking for is an ontology able to grow and not a static adaptation of some knowledge representation.

One important aspect is that most successful solutions integrate different resources for retrieving information and also as reference knowledge for detecting wrong alignments.

Most methods offer automation of only some steps of the generation process. Modular solutions, rather then monolithic applications should offer a better architecture for covering the larger part of the ontology life cycle, although integration of steps is mostly manual.

In order to be able to fulfill our goal, there still remains a lot of work that we could divide into three main actions: i) one is the automatic construction of a dynamic reference ontology; ii) the second is to build applications able to integrate this new approach (that we could call *semantic method* in opposition to the *exactness method*) and further investigating the new types of exceptions that it could involve; iii) and more in the semantic web area to further develop a new methodology for automatic ontology generation.

# Références

## Résumé

Dans cet article, nous présentons une méthode et un outil pour dériver le noyau d'une ontologie à partir de plusieurs schémas XML. Nous rappelons d'abord ce qu'est une ontologie et ses rapports avec les schémas XML. Ensuite, nous nous concentrons sur la méthodologie de construction de l'ontologie et les exigences pour les outils associés. Ensuite, nous introduisons Janus, un outil pour construire une ontologie à partir de schémas XML différents mais dans un même domaine. Nous résumons les caractéristiques principales de Janus et illustrons ses fonctionnalités sur un exemple simple. Finalement, nous comparons notre outil avec ceux existants pour aider à la construction d'ontologie.